\documentclass[seceq]{ptptex}

\usepackage{graphicx}
\usepackage{wrapft}
\usepackage{bm}% bold math
\def\Vec#1{\mbox{\boldmath $#1$}}

\def\itmb{\begin{itemize}}
\def\itme{\end{itemize}}
\def\enmb{\begin{enumerate}}
\def\enme{\end{enumerate}}
\def\eqnb{\begin{equation}}
\def\eqne{\end{equation}}

\def\NPB{{Nucl. Phys.} { B}}
\def\PLB{{Phys. Lett.} B}

\def\PRD{{Phys. Rev.} D}

\title{%        %You can use \\ for explicit line-break
The Color Antisymmetric Ghost Propagator \\ and One-Loop Vertex Renormalization%
}

\author{%       %Use \scshape  for the family name
Sadataka \textsc{Furui}%
}

\inst{%     %Affiliation, neglected when [addenda] or [errata]
School of Science and Engineering, Teikyo University\\
 Utsunomiya 320-8551, Japan
}

\recdate{September 19, 2007}%            %Editorial Office will fill in this.

\abst{%         %this abstract is neglected when [addenda] or [errata]
The color matrix elements of the ghost triangle diagram that appears in the 
triple gluon vertex and the ghost-ghost-gluon triangle diagram that
appears in the ghost-gluon-ghost vertex are calculated. 
The ghost-ghost-gluon triangle contains
a loop consisting of two color diagonal ghosts and one gluon and a loop
consisting of two color antisymmetric ghosts and one gluon.  Consequently,
the pQCD argument in the infrared region based on the one particle irreducible
diagram should be modified.

Implications for the Kugo-Ojima color confinement and the QCD running coupling
are discussed.
}

\begin{document}
\newcommand{\ttbs}{\char'134}
\newcommand{\Slash}[1]{\ooalign{\hfil/\hfil\crcr$#1$}}

\maketitle

\section{Introduction}
In 1971, 't Hooft\cite{tH71} showed that in massless Yang-Mills field theory, non-gauge
invariant regulator fields can be incorporated, provided that in the limit of high regulator mass the gauge invariance can be restored by means of a finite number of 
counter terms including ghost fields and longitudinally polarized gauge fields. 
After this work, Taylor\cite{Tay71} pointed out that if the charge carried by
the ghost and the gauge field are the same, the Ward identity in QED, 
i.e. $Z_1/Z_2=1$, where $Z_1$ is the vertex renormalization factor and $Z_2$ is the matter field wave function renormalization factor, can be extended to QCD as $Z_1/Z_3=Z_{\bar\psi\psi A}/Z_\psi$, where $Z_3$ is the gluon wave function renormalization factor, $Z_\psi$ is the matter field wave function renormalization factor, and $Z_{\bar\psi\psi A}$ is the vertex renormalization factor of the matter field.

In these arguments, vertices at tree level are considered, and the ghost propagators were assumed to be color diagonal.  
The ghost propagator is defined as the Fourier transform (FT) of the matrix element of the inverse Faddeev-Popov operator,

\begin{eqnarray}
FT[D_G^{ab}(x,y)]&=&FT\langle {\rm tr} ( \Lambda^a \{({\cal  M})^{-1}\}_{xy}
\Lambda^b ) \rangle\nonumber\\
&=&\delta^{ab}D_G(q^2),  \nonumber
\end{eqnarray}
where ${\mathcal M}=-\partial_\mu D_\mu$, and $\{ \}_{xy}$ represents the matrix value.
We define the ghost dressing function $G(q^2)$ as $q^2 D_G(q^2)$. In the Dyson-Schwinger (DS) approach, $G(q^2)$ at 0 momentum behaves as $\sim (q^2)^{-\kappa}$, and that of the gluon dressing function $Z(q^2)$ behaves as $\sim (q^2)^{2\kappa}$\cite{AS00}.

In lattice simulations\cite{FN04,FN06a, FN06b, FN06d}, we calculated the overlap to obtain the color diagonal ghost propagator,
\begin{equation}
D_G(q)=\frac{1}{N_c^2-1}\frac{1}{V}{\rm tr}\left\langle \delta^{ab}(\langle \Lambda^a\cos{ q}\cdot{ x}|f_c^b({ x})\rangle+\langle \Lambda^a\sin{ q}\cdot{ x}|f_s^b({ x})\rangle)\right\rangle, \nonumber
\end{equation}
and the color antisymmetric ghost propagator
\begin{equation}
\phi^c(q)=\frac{1}{\mathcal N}\frac{1}{V}{\rm tr}\left\langle f^{abc}(\langle \Lambda^a\cos{ q}\cdot{ x}|f_s^b({ x})\rangle-\langle \Lambda^a\sin{ q}\cdot{ x}|f_c^b({ x})\rangle)\right\rangle,\nonumber
\end{equation}
where ${\mathcal N}=2$ for $SU(2)$ and 6 for $SU(3)$. Here, ${f_c}^b({ x})$ and ${f_s}^b({ x})$ represent the solution $f^b({ x})={\mathcal M[U]}^{-1} \rho^b({ x})$, where $U$ is the lattice link variable, with $\displaystyle \rho_c^b({ x})=\frac{1}{\sqrt V}\Lambda^b\cos{q\cdot x}$ and $\displaystyle \rho_s^b({ x})=\frac{1}{\sqrt V}\Lambda^b\sin{ q\cdot x}$, respectively. We normalize the $SU(3)$ generator $\Lambda^a$ as
$\displaystyle \Lambda^a=\frac{\lambda}{\sqrt 2}$, where $\lambda$ is that defined by Gell-Mann. In our notation, we have ${\rm tr}\Lambda^a\Lambda^b=\delta^{ab}$.

Lattice studies of the color antisymmetric ghost propagator in quenched $SU(2)$\cite{CMM04} and unquenched $SU(3)$\cite{FN06a, FN06b, FN06d} were motivated by  analysis in the local composite operator (LCO) approach based on symmetry under the BRST (Becchi, Rouet, Stora and Tyutin) transformation\cite{Du03,LSS02}.
The presence of $\langle f^{abc}\bar c^b c^c\rangle$ condensates was expected to
manifest itself in the color anti-symmetric ghost propagator.
The ghost condensates, as the on-shell BRST partner of $A^2$ condensates, were also discussed in Ref.\cite{Ko01}. In the study of unquenched $SU(3)$\cite{FN06a, FN06b, FN06d}, 
the modulus of the color antisymmetric ghost propagator is found to be relatively large, and its infrared singularity, characterized by $\alpha_G'\sim 0.9$, is larger than that of the color diagonal ghost propagator $\alpha_G\sim 0.25$\cite{FN06a}.
 However, analysis of quenched $SU(3)$ on a $56^4$ lattice\cite{FN06d} showed that the modulus of the color antisymmetric ghost propagator is small and its sample variation is large.  

These data suggest that the ghost propagators of $SU(2)$ and $SU(3)$ are qualitatively different, although the gluon propagators of the two are similar, and that
the presence of the dynamical quark affects the color antisymmetric ghost propagator.

In the momentum subtraction scheme ($\widetilde{MOM}$ scheme),  the running couplings in the Coulomb gauge and in the Landau gauge are calculated as the product of color diagonal ghost dressing function squared and the gluon propagator. 
The running coupling in the Landau gauge is
\[
\alpha_s(q)=q^6 D_G(q)^2 D_A(q),
\]
where $D_A(q)$ is the gluon propagator in 4-dimensional space,
and that in the Coulomb gauge is 
\[
\alpha_I({\Vec q})={\Vec q}^5 D_G({\Vec q})^2 {D_A}^{tr}({\Vec q}),
\]
where ${D_A}^{tr}({\Vec q})$ is the 3-dimensional transverse
gluon propagator.  

In the Coulomb gauge, the color-Coulomb potential defines another running coupling $\alpha_{Coul}({\Vec q})=\frac{11N_c-2N_f}{12N_c}q^2V_{Coul}({\Vec q})$. Because we do not fix $A_0$,  the variation of $\alpha_{Coul}({\Vec q})$ in the infrared region is large.

We observed that the running coupling
in the Landau gauge, $\alpha_s(q^2)$, has a peak around 2.3, near $q=0.4$ GeV, but it is strongly suppressed below 0.4 GeV, whereas that in the Coulomb gauge $\alpha_I({\Vec q})$ exhibits freezing to a constant around 3\cite{FN07}.
This qualitative difference is believed to be due to the difference between the
ghost propagators: That in the Coulomb gauge is 3 dimensional and instantaneous, while that in the Landau gauge is 4 dimensional and propagating in the 4th direction.

In usual perturbative QCD (pQCD), a ghost is assumed to be color diagonal, which is valid at high energy.  In a study of the Gribov horizon, Zwanziger introduced a real Bose (ghost) field $\phi^{ab}_\mu(x)$ and the additional action\cite{Zw89,Zw94}
\begin{eqnarray}
S&=&S_{cl}+\gamma S_1+S_2 \nonumber\\
S_{cl}&=&\frac{1}{4}\sum_{\mu,\nu,a}\int d^4 x(F_{\mu\nu}^a(x))^2\nonumber\\
S_1&=&-(2C)^{-1}\int\int d^4 x d^4 y \quad tr A_\mu(x){\mathcal M}[A]^{-1}A_\nu (y)\nonumber\\
S_2&=&\int d^4 x\left(\frac{1}{2}\phi^{ab}_\mu \mathcal M^{bd}[A]\phi^{ad}_\mu+i\gamma^{1/2}C^{-1/2}f^{abd}\phi^{ad}_\mu {A_\mu}^b\right),
\end{eqnarray}
in order to restrict the gauge configuration in the fundamental modular region.
Here, $C$ is the value of the $SU(3)$ Casimir operator of the fundamental representation.

In the Gribov-Zwanziger Lagrangian, the equation of motion for $\phi^{ad}_\mu$ becomes\cite{Gr06}
\begin{equation}
(D_\mu\phi_\nu)^{ab}=\partial_\mu{\phi_\nu}^{ab} f^{acd}A^c_\mu{\phi_\nu}^{db},
\end{equation}
and its solution is
\begin{equation}
{\phi_\mu}^{ab}=-\frac{\gamma^2}{\sqrt 2}f^{abc}{\mathcal M[A]}^{-1}{A_\mu}^c.
\end{equation}
When ${A_\mu}^c$ is replaced by $q_\mu c^c$, where $c^c$ is the ghost field and 
$f^{deh}$ is multiplied, we obtain
\begin{equation}
f^{deh}\phi^{ab}_\mu =-\frac{\gamma^2}{\sqrt 2}(\delta^{ae}\delta^{bh}-\delta^{be}\delta^{ah}){\mathcal M[A]}^{-1}q_\mu c^c.
\end{equation}

The color antisymmetric ghost, $\phi^c(q)$, multiplied by the momentum $q_\mu$ 
and $f^{deh}$ is  similar to Zwanziger's ghost, $\phi^{ab}_\mu$.  In an analytical one-loop
calculation, based on the Gribov-Zwanziger Lagrangian, including two additional ghost field denoted by $\{\phi^{ab}_\mu, \bar\phi^{ab}_\mu\}$ and $\{\omega^{ab}_\mu,\bar\omega^{ab}_\mu\}$, 
infrared freezing of the running coupling $\alpha_s(q)$ was demonstrated\cite{Gr06}. 
Although Zwanziger's ghost, $\phi^{ab}_\mu$, was introduced in order to restrict the gauge configuration into the fundamental modular region, its relation to the $A^2$ condensates is discussed in Ref.\cite{Du05}. 
Although the color antisymmetric
ghost multiplied by the momentum $q_\mu$ and $f^{deh}$ and Zwanziger's ghost are different, we expect similar corrections to the infrared exponents of the gluon propagator and the ghost propagator 
through the color antisymmetric ghost propagator in the Landau gauge, when
it is incorporated properly.

The organization of this paper is as follows. In $\S$ 2 we study the triple gluon vertex
and the ghost-gluon-ghost vertex in one loop.
In $\S$ 3, the contribution of the ghost loop in the gluon propagator is
investigated, and in $\S$ 4, the contribution of the gluon-ghost-ghost loop
in the ghost propagator is studied. Discussion of the quark-gluon vertex, the Kugo-Ojima confinement parameter and QCD running coupling is given in $\S\S$ 5 and 6. A conclusion and discussion on the outlook are given in $\S$ 7.

\section{The triple gluon vertex and the ghost-gluon-ghost vertex}
In this section we study the contribution of the color antisymmetric ghost
propagator in the triple gluon vertex and the ghost-gluon-ghost vertex. 

\begin{figure}[htb]
\parbox{\halftext}{
\centerline{\includegraphics[width=4 cm, height=5 cm]{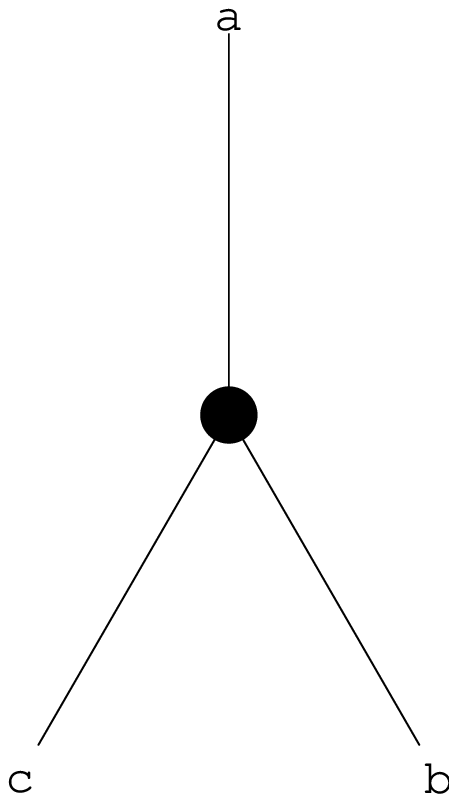}}
\caption{The triple gluon vertex.  }}
\label{gv}
\hfill
\parbox{\halftext}{
\centerline{\includegraphics[width=4 cm, height=5 cm]{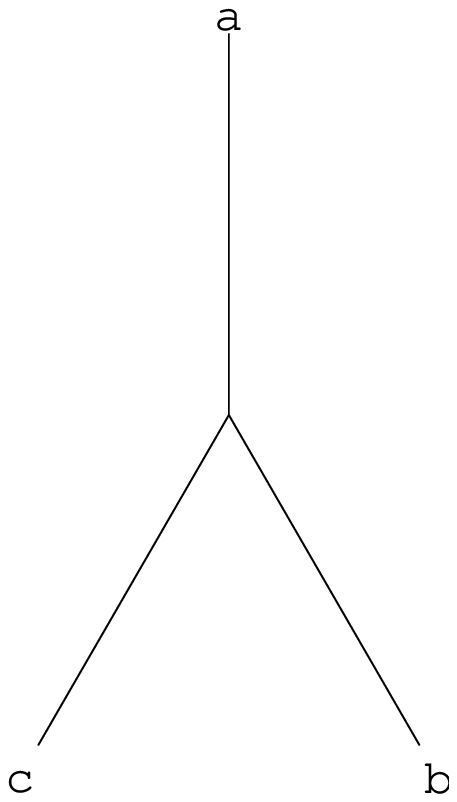}}
\caption{The bare triple gluon vertex. }}
\label{gv1}
\end{figure}

The triple gluon vertex defined in Fig. 1 %\ref{gv} 
in the pQCD is given by the tree-level diagram (Fig. \ref{gv1}), ghost loop diagram (Fig. 3) %(\ref{gv2})
 and the gluon loop diagram (Fig. 4) %\ref{gv4_n}).
\begin{figure}[htb]
\parbox{\halftext}{
\centerline{\includegraphics[width=5 cm, height=6 cm]{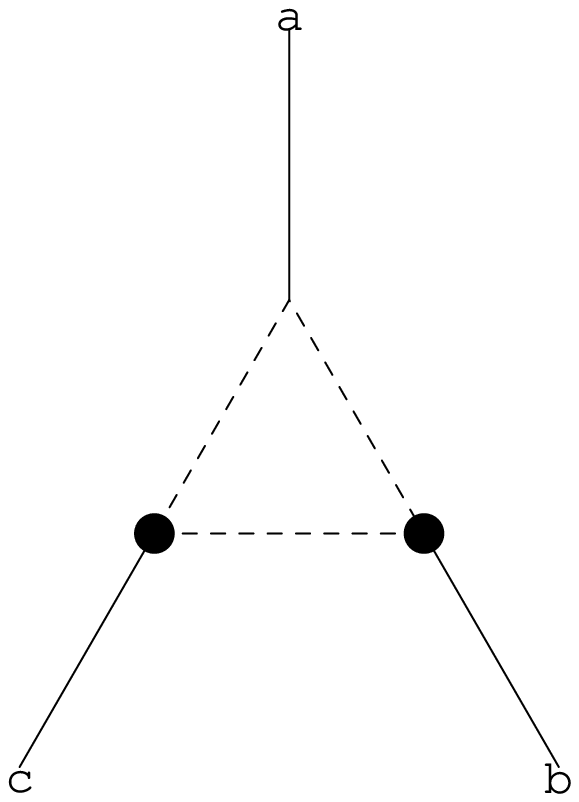}}
\caption{The dressed triple gluon vertex. The dashed line represents a ghost and the thin line a gluon.}}
\label{gv2}
\hfill
\parbox{\halftext}{
\centerline{\includegraphics[width=5 cm, height=6 cm]{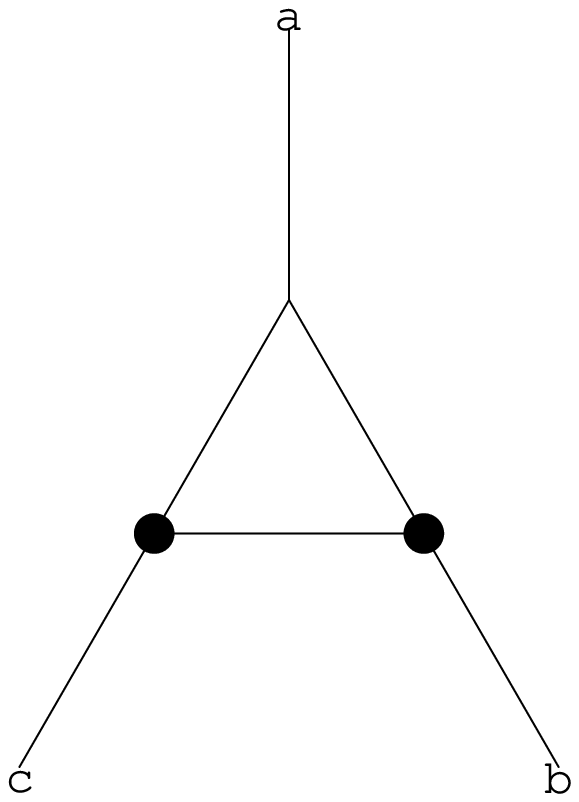}}
\caption{The triple gluon vertex. The  thin lines represent gluons.}
\label{gv4_n}}
\end{figure}

\begin{figure}[htb]
\centerline{\includegraphics[width=5 cm, height=6 cm]{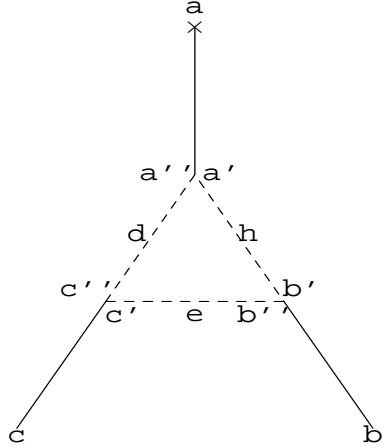}}
\caption{The triple gluon vertex. The dashed line represents a ghost and the thin line is a gluon.}
\label{gv3}
\end{figure}

The color indices of the ghost loop in the triple gluon vertex
are assigned as in Fig. \ref{gv3}.
We express the ghost propagator, which is assigned as $d$, as a combination of the
color diagonal and color antisymmetric pieces:
\begin{equation}
\delta^{a''c''}D_G(k)+2f^{a''c''d}\phi_d(k).
\end{equation}
The same combination is assumed for the ghost assigned as $h$.
The color factor
\begin{equation}
f^{a''a'a}f^{b''b'b}f^{c''c'c}
\end{equation}
is multiplied at the three edges of the triangle.

%The contribution of the triple gluon loop in the triple gluon vertex
%is given in Fig.\ref{gv4_n}.

The ghost-gluon-ghost vertex is expressed as in Fig. 6 %\ref{ghv} 
and its tree-level diagram is given in Fig. 7. %\ref{ghv1}
At the one-loop level, the ghost-ghost-gluon loop shown in Fig. 8 %\ref{ghv2} 
contributes.  The assignment of the color indices is shown in Fig. 9. %\ref{ghv3}
\begin{figure}[htb]
\parbox{\halftext}{
\centerline{\includegraphics[width=4 cm, height=5 cm] {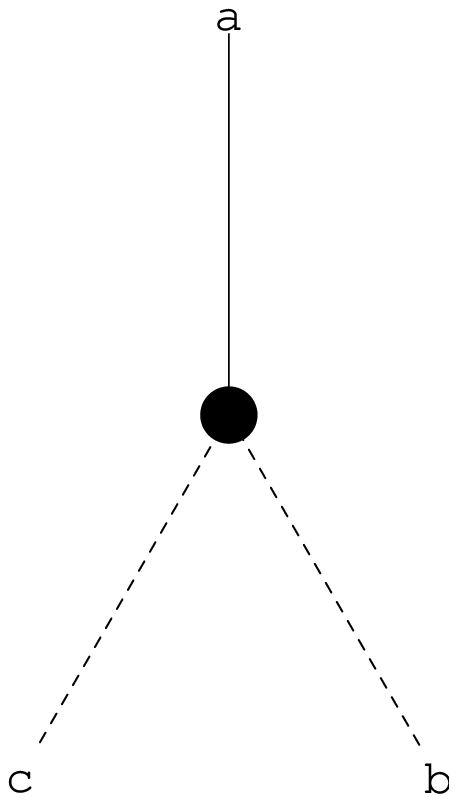}}
\caption{The gluon-ghost-ghost vertex. The dashed line represents a ghost and the thin line a gluon.}}
\label{ghv}
\hfill
\parbox{\halftext}{
\centerline{\includegraphics[width=4 cm, height=5 cm]{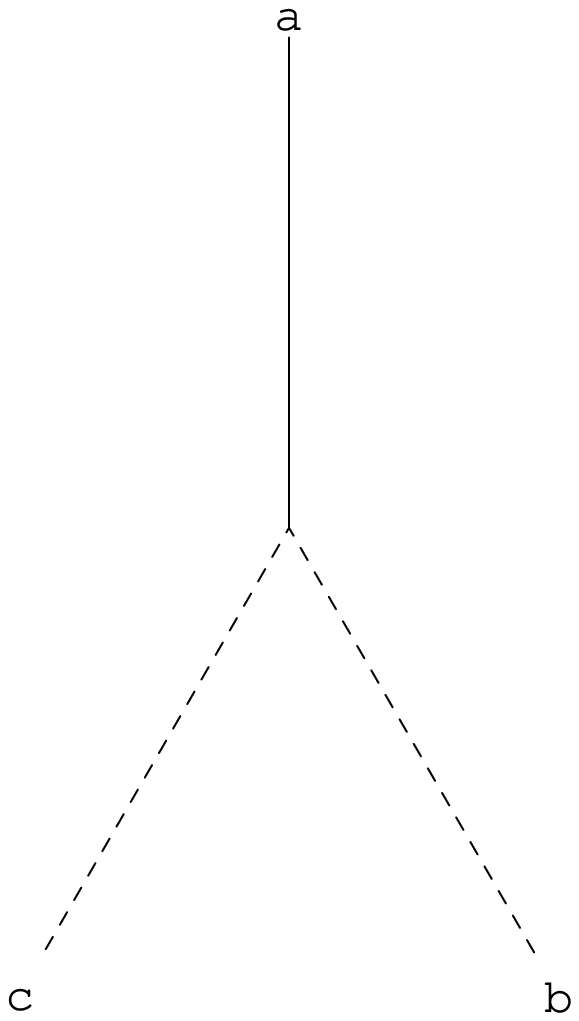}}
\caption{The gluon-ghost-ghost vertex. The dashed line represents a ghost and the thin line a gluon.}}
\label{ghv1}
\end{figure}
\begin{figure}[htb]
\parbox{\halftext}{
\centerline{\includegraphics[width=5 cm, height=6 cm] {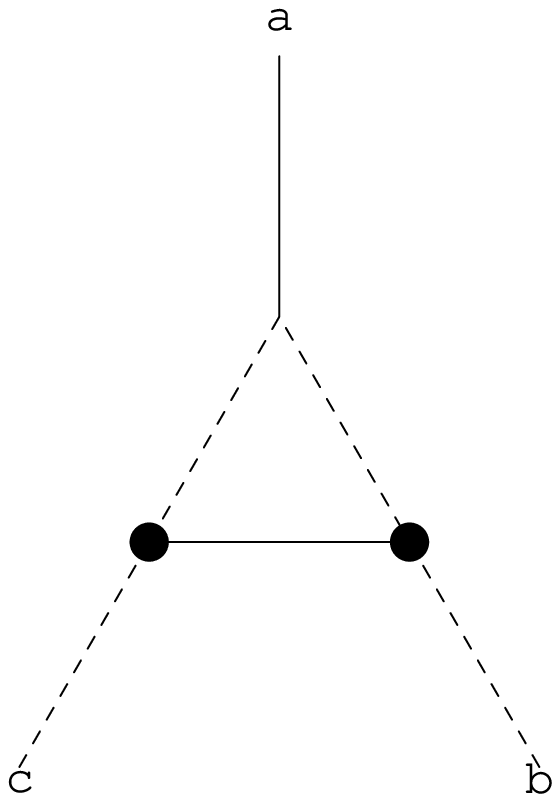}}
\caption{The gluon-ghost-ghost vertex. The dashed line represents a ghost and the thin line a gluon.}}
\label{ghv2}
\hfill
\parbox{\halftext}{
\centerline{\includegraphics[width=5 cm, height=6 cm]{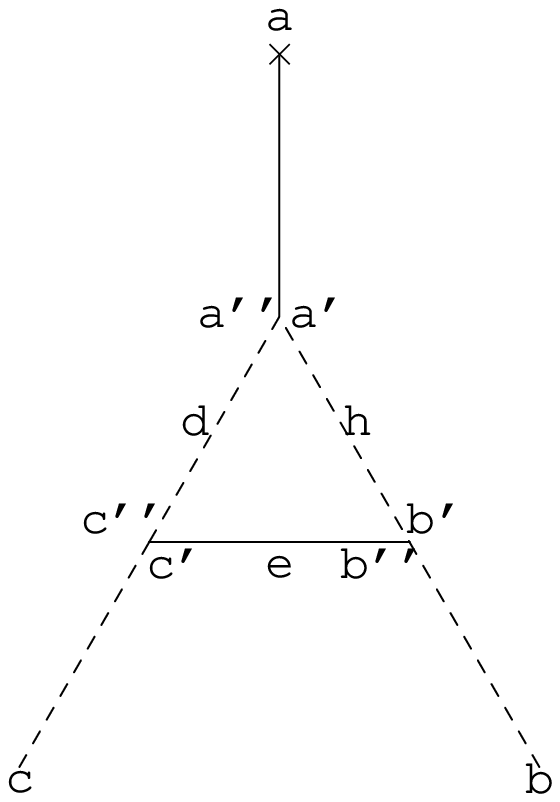}}
\caption{The gluon-ghost-ghost vertex. The dashed line represents a ghost and the thin line a gluon.}}
\label{ghv3}
\end{figure}

In the calculation of the color matrix element of the one loop vertex diagram
in Figs. \ref{gv3} and 9, %\ref{ghv3}, 
we fix $a,b,c$ and carry out a summation over the color indices $a',a'',b',b'',c',c'',d,e,h$.
The matrix elements of the quark gluon vertices are proportional to $f^{abc}$,
which are given by
\begin{itemize}
\item $f^{123}=1$,
\item $\displaystyle f^{147}=f^{246}=f^{345}=f^{516}=f^{257}=f^{637}=\frac{1}{2}$, 
\item $\displaystyle f^{458}=f^{678}=\frac{\sqrt 3}{2}$.
\end{itemize}

Using the definitions $D_{c'b''}\equiv D_e$,
$D_{a'b'}\equiv D_h$ and 
 $D_{a''c''}\equiv D_d$, we obtain
in the case of $SU(2)$ the coefficients given in Table \ref{tabsu2}.
In this table, ${abc}$ and ${deh}$ are the $SU(2)$ color indices, $D_hD_dD_e$ represents
all color diagonal $D_h\phi_d\phi_e$, and $D_e\phi_h\phi_d$ represent one propagator
assigned as $D_h$ or $D_e$ are color diagonal but the rest are color antisymmetric.

When $deh$ are not elements of the Cartan subalgebra, there appear terms
proportional to $\phi\phi\phi$, but the coefficient is purely imaginary,
and both plus sign terms and minus sign terms appear, and they cancel.

\begin{table}[htb]
\begin{center}
\begin{tabular}{cccccccc}
  $abc$ & $deh$ & $D_hD_dD_e$ &$D_h\phi_d\phi_e$  &$D_d\phi_e\phi_h$& $D_e\phi_h\phi_d$  \\
\hline
 123 & 333 & -1 & -4 & 0 & 0 \\
\hline
\end{tabular}
\caption{The $SU(2)$ color matrix elements of the ghost triangle diagram.} \label{tabsu2}
\end{center}
\end{table}

In the case of $SU(3)$, we obtain the coefficients which are to multiply $f^{abc}$, as in Table \ref{tabsu3II}. The gluon that has the color index in the Cartan subalgebra couples with a pair of color antisymmetric ghost propagators and the contribution has the same sign as the contribution of the gluon that couples with a pair of color diagonal ghost propagators. 

The qualitative difference between the color matrix elements of $SU(2)$ and $SU(3)$ comes
from the anomaly cancelling equation of the triangle diagram
\begin{equation}
{\rm tr}\{\{\Lambda^a,\Lambda^b\}\Lambda^c\}=0,
\end{equation}
where $\{ \}$ represents the anti-commutator, is satisfied for $SU(2)$ but not for $SU(3)$.

\begin{table}[htb]
\begin{center}
\begin{tabular}{c c c c c c c}
  $abc$ & $deh$ & $D_hD_dD_e$ &$D_h\phi_d\phi_e$  &$D_d\phi_e\phi_h$& $D_e\phi_h\phi_d$  \\
\hline
 123 & 888 & -1.5 & -1.5 & 1.5 & 1.5 \\
     & 333 & -1.5 & -4.5 & 0   &   0 \\
     & 833 & -1.5 &  0   & -0.5 & 0 \\
     & 383 & -1.5 &  0   & 0  & -0.5 \\
     & 338 & -1.5 &  -4.5 & 0 & 0 \\
     & 838 & -1.5 & 0   &  0 & 1.5 \\
     & 883 & -1.5 & -1.5 & 0 & 0 \\
     & 388 & -1.5 & 0 & 1.5 & 0 \\
\hline
 147 & 888 & -1.5 & 0 & 0 & 1.5 \\
     & 333 & -1.5 & -1 & 1 & -0.5 \\
     & 338 & -1.5 & -1 & 0 & $-\sqrt 3/2$\\
     & 383 & -1.5 & $-\sqrt 3$ & $-\sqrt 3$ & -0.5 \\
     & 833 & -1.5 & 0 & 1 & $\sqrt 3/2$ \\
     & 883 & -1.5 & 0 & $-\sqrt 3$ & $\sqrt3/2$ \\
     & 838 & -1.5 & 0 & 0 & 1.5 \\
     & 388 & -1.5 & $-\sqrt 3$& 0 & $\sqrt 3/2$ \\
\hline
 458 & 888 & -1.5 & 4.5 & 0 & 0 \\
     & 333 & -1.5 & -1.5 & 1 & 1\\
     & 338 & -1.5 & -1.5 & 0 & 0\\
     & 383 & -1.5 & $-\sqrt 3/2$ & $-\sqrt 3$ & 1\\
     & 833 & -1.5 & $-\sqrt 3/2$ & 1 & $-\sqrt 3$\\ 
     & 388 & -1.5 & $-\sqrt 3/2$ & 0 & 0\\
     & 883 & -1.5 & -4.5 & $-\sqrt 3$ & $-\sqrt 3$\\
     & 838 & -1.5 & $-\sqrt 3/2$ & 0 & 0\\
\hline
 453 & 888 & -1.5 & -1.5 & 0& 0 \\
     & 333 & -1.5 & -4.5 & 1  &   1 \\
     & 833 & -1.5 & $\frac{3\sqrt 3}{2}$  & 1 & $\sqrt 3$ \\
     & 383 & -1.5 & $-\frac{3\sqrt 3}{2}$  & $\sqrt 3$  & 1 \\
     & 338 & -1.5 &  -4.5 & $-2\sqrt 3$ & $-2\sqrt 3$ \\
     & 838 & -1.5 & $\frac{3\sqrt 3}{2}$   & $-2\sqrt 3$& 0 \\
     & 883 & -1.5 & -1.5 & $\sqrt 3$ & $\sqrt 3$ \\
     & 388 & -1.5 & 0 & $-\frac{3\sqrt 3}{2}$ & $-2\sqrt 3$\\
\hline
 678 & 888 & -1.5 & -4.5 & 0 & 0 \\
     & 333 & -1.5 & -1.5 & 1 & 1\\
     & 338 & -1.5 & -1.5 & 0 & 0\\
     & 383 & -1.5 & $\sqrt 3/2$ & $\sqrt 3$ & 1\\
     & 833 & -1.5 & $\sqrt 3/2$ & 1 & $\sqrt 3$\\ 
     & 388 & -1.5 & $\sqrt 3/2$ & 0 & 0\\
     & 883 & -1.5 & -4.5 & $\sqrt 3$ & $\sqrt 3$\\
     & 838 & -1.5 & $\sqrt 3/2$ & 0 & 0\\
\hline 
\end{tabular}
\caption{The $SU(3)$ color matrix elements of the ghost triangle.}\label{colormatrix2}\label{tabsu3II}
\end{center}
\end{table}

The coefficients of the $D\phi\phi$ terms of $f^{458}$ and $f^{678}$ differ only in their signs. 
When the color indices are $deh$ = 333 and $abc$ = 123, the same structure as in the $SU(2)$ appears. But in the case of $deh$ = 888 and $abc$ = 147 since $f^{458}$ and $f^{678}$ are the only coefficient that do not vanish when coupled to the color source 8, the
color antisymmetric pair of $d$ and $e$ appear in the link.

In the case of the unquenched $SU(3)$ configuration with the Kogut-Susskind fermion MILC$_f$, the ratio of the dressing function  $q^2\phi(q)/G(q)$ for $q\sim 0.18$ GeV is about 0.2 and for $q\sim 0.4$GeV is 0.08\cite{FN06b,FN06d}, while in the case of quenched $SU(2)$, the ratio for $q\sim 0.25$GeV is about 0.1 \cite{CMM04,BCLM02}and for $q\sim 0.4$ GeV is $0.05-0.06$\cite{FN06a,CMM04,BCLM02}. 

\section{The ghost loop in the gluon propagator}
\begin{figure}[tb]
\parbox{\halftext}{
\centerline{\includegraphics[width=5 cm, height=1.5 cm]{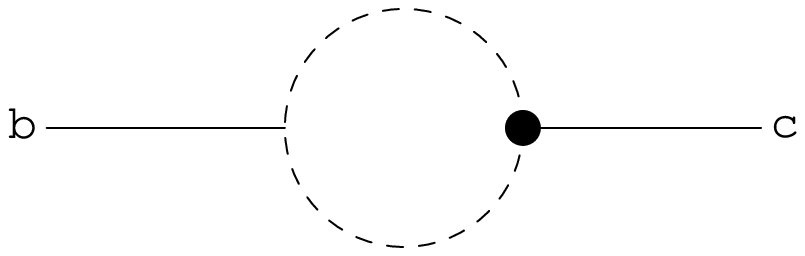}}
\caption{The gluon propagator dressed by the ghost propagator. }
\label{gl_prop1}}
\hfill
\parbox{\halftext}{
\centerline{\includegraphics[width=5 cm, height=1.5 cm]{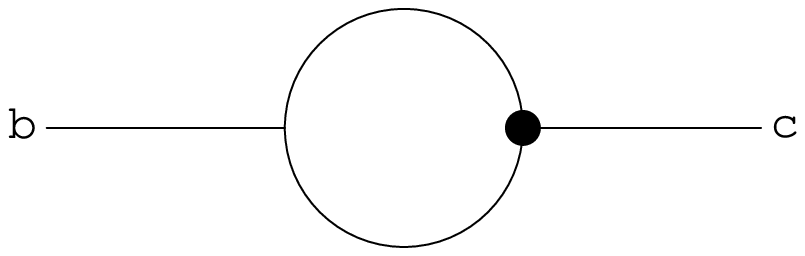}}
\caption{The gluon propagator dressed by gluon propagator.}
\label{gl_prop2}}
%\end{minipage}
\end{figure}

At one-loop order, the vacuum polarization tensor that contributes to the gluon 
self-energy consists of a) a quark loop, b) a ghost loop (Fig. \ref{gl_prop1}), c) a gluon tadpole and d) a gluon loop (Fig. \ref{gl_prop2}). In  b) there is a ghost-gluon-ghost vertex, and in d) there is a triple gluon vertex at two-loop order, in which the color antisymmetric ghost could contribute.
In this section, we study the effect of the ghost loop on the gluon propagator.

The product of the color antisymmetric ghost propagator contributes to the 
ghost-loop diagram as shown in Fig. 12. %\ref{tadpole1}. 
There the cross indicates a coupling to the quark loop, as shown in Fig. 13, %\ref{tadpole2} 
or to gluon loops. The quark loop or the gluon loop can couple to other gluons, 
and among the ghost propagator specified by $d,e$ and $h$, one is
color diagonal and the other two are color antisymmetric. The color
matrix elements we calculated in the triple gluon vertex imply that
  if the ghost propagator that does not couple with the gluon of color
index $a$ is color diagonal and the rest are color antisymmetric, the color index 
$a$ should belong to the Cartan subalgebra, in order that the matrix element becomes real.

\begin{figure}[tb]
\parbox{\halftext}{
\centerline{\includegraphics[width=5 cm, height=2.4 cm]{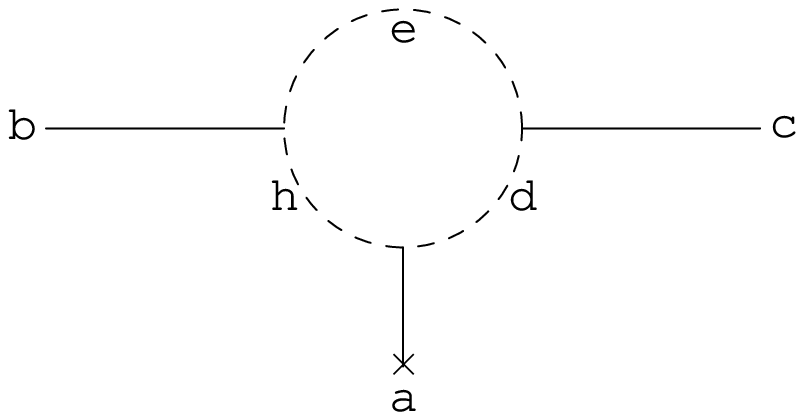}}
\caption{The ghost loop contribution in the gluon propagator. The dashed line represents a ghost. The $\times$ indicates the dressing of a gluon whose color index is $a$.}}
\label{tadpole1}
\hfill
\parbox{\halftext}{
\centerline{\includegraphics[width=5 cm, height=3.5 cm]{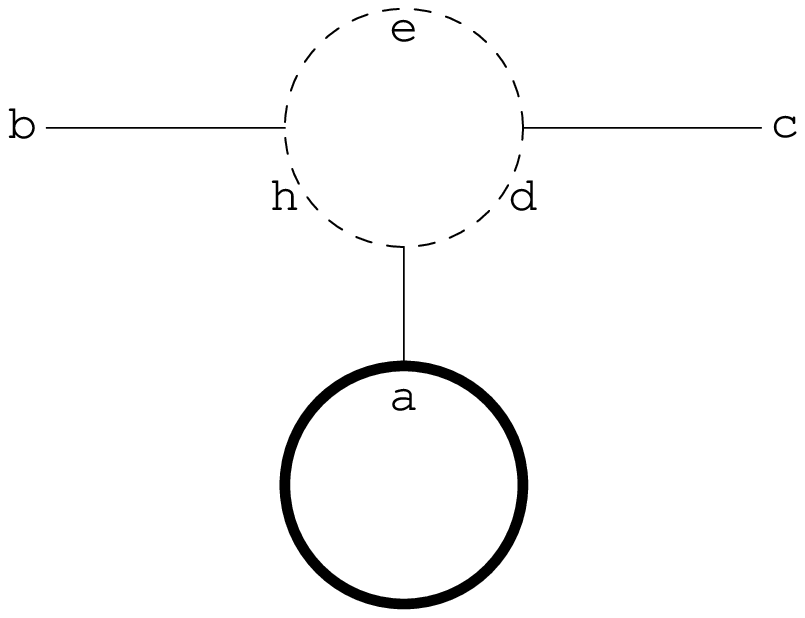}}
\caption{The ghost loop contribution in the gluon propagator. The dashed line represents a ghost and the thick line is a quark.}}
\label{tadpole2}
\end{figure}

 But because other gluons can couple to the quark loop, the color index $a$ is
not necessarily in the Cartan subalgebra.   The propagator is proportional to
$f^{abc}$, and since 
the signs of the color matrix element given in Table\ref{tabsu3II} are random, its 
contribution is expected to be small, and thus the color mixing of gluons should
be small. Lattice simulations also indicate that the gluon propagator is
 diagonal in the color.

\section{The ghost-ghost-gluon loop in the ghost propagator}
At one-loop order, the ghost self-energy is given by the gluon-ghost loop 
shown in Fig. 14. %\ref{gh_prop1}

At two-loop order, the ghost-ghost-gluon loop contributes in such a manner that
a pair of color antisymmetric ghost propagator and the gluon propagator are
incorporated.
We consider the production of the color antisymmetric ghost pair 
from a gluon of color index $a$. The propagator shown in Fig. 15 %\ref{gh_prop1}
 is proportional to $f^{abc}$, and its trace vanishes. 
However, the product of these propagators shown in Fig. 16 %\ref{ghgh_propg} 
does not vanish, and they contribute to the dressing of the external ghosts of the ghost-gluon-ghost vertex.

\begin{figure}
\parbox{\halftext}{
\centerline{\includegraphics[width=5 cm, height=1.5 cm]{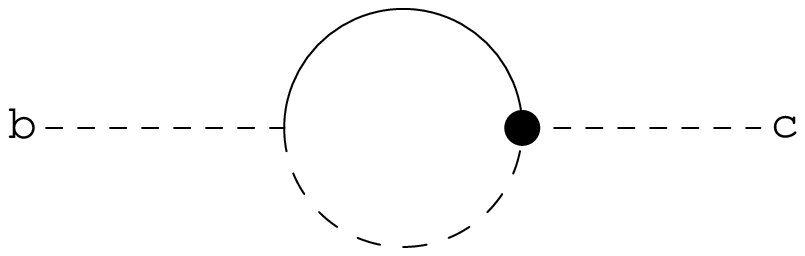}}
\caption{The ghost propagator.}}
\label{gh_prop1}
\hfill
\parbox{\halftext}{
\centerline{\includegraphics[width=5 cm, height=4 cm]{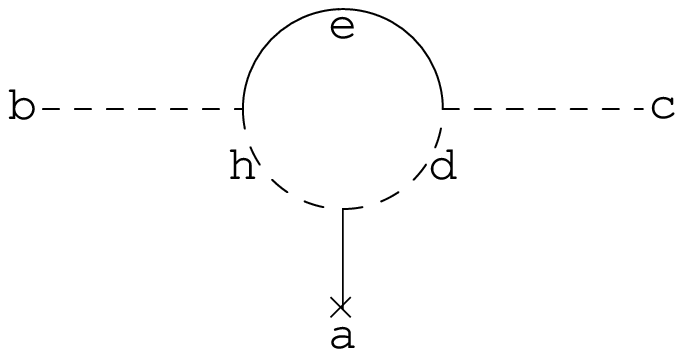}}%
\caption{The dressing of the ghost propagator by the gluon. The dashed line is a ghost the thin line is a gluon. The $\times$ indicates dressing of a gluon in Cartan subalgebra.}}
\label{gh_propg}
\end{figure}

When the gluon with color index $a$ can be treated as a background field that couples to a quark loop, as shown in Fig. 17 %\ref{gh_dress_tl}, 
the color antisymmetric ghost propagator contributes.
 In contrast to the gluon propagator, the color mixing of the ghost is not ruled out from the lattice simulations.

\begin{figure}[htb]
\parbox{\halftext}{
\centerline{\includegraphics[width=5 cm, height=4.5 cm]{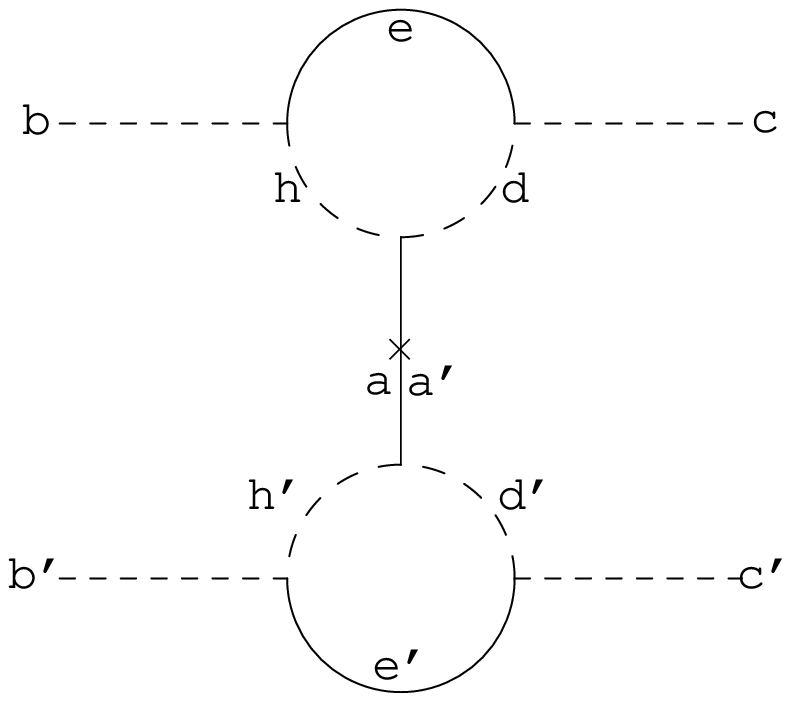}}
\caption{The two-ghost propagator. The $\times$ indicates a gluon of color indices $a=a'$ in the Cartan subalgebra.}}
\label{ghgh_propg}
\hfill
\parbox{\halftext}{
\centerline{\includegraphics[width=5 cm, height=4 cm]{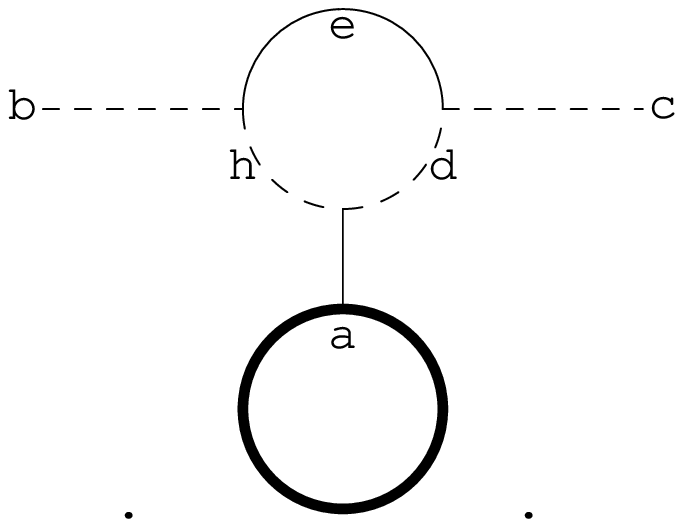}} 
\caption{The dressing of the ghost propagator by the gluon. The dashed line represents a ghost the thin line a gluon, and the thick line a quark.}}%\protect\\}}
\label{gh_dress_tl}
\end{figure}

The dressing of the ghost propagator by the gluon is depicted in Fig. 14. % \ref{gh_propg}. 
The integral is given by
\begin{equation}
d(b,c)=\int\frac{d^4 q}{(2\pi)^4} \delta^{bc}\left(\frac{\delta_{\mu\nu}q_\mu q_\nu}{(p-q)^2 q^{2(1+\kappa)}}-\frac{q_\mu q_\nu(p-q)_\mu(p-q)_\nu}{(p-q)^4 q^{2(1+\kappa)}}\right)Z_3((p-q)^2).
\end{equation}
When $Z_3((p-q)^2)$ is taken as a constant, the formulae in Ref.\cite{AFRvS03} give\begin{eqnarray}
d(b,c)&\propto& K_2(1+\kappa,1,p)p^2-(K_2(1+\kappa,2,p)p^4-2 L_2(1+\kappa,2,p)p^4\nonumber\\
&&+M_3(1+\kappa,2,p)p^4),
\end{eqnarray}
where
\begin{eqnarray}
&&\int d^4 q \frac{q_\mu q_\nu}{q^{2a}(P-q)^{2b}}=K_1(a,b,p)p_\mu p_\nu +K_2(a,b,p)p^2\delta_{\mu\nu},\nonumber\\
&&\int d^4 q\frac{q_\mu q_\nu q_\rho}{q^{2a}(p-q)^{2b}}=L_1(a,b,p)p_\mu p_\nu p_\rho%\nonumber\\
+L_2(a,b,p)p^2(p_\mu\delta_{\nu\rho}+p_\nu\delta_{\rho\mu}+p_\rho\delta_{\mu\nu}),\nonumber\\
&&\int d^4 q \frac{q_\mu q_\nu q_\rho q_\sigma}{q^{2a}(p-q)^{2b}}=M_1(a,b,p)p_\mu p_\nu p_\rho p_\sigma\nonumber\\
&&+M_2(a,b,p)p^2(\delta_{\mu\nu}p_\rho p_\sigma+\delta_{\mu\rho}p_\nu p_\sigma
+\delta_{\mu\sigma}p_\rho p_\mu+\delta_{\nu\rho}p_\mu p_\sigma+\delta_{\nu\sigma}p_\rho p_\mu+\delta_{\rho\sigma}p_\mu p_\nu) \nonumber\\
&&+M_3(a,b,p)p^4(\delta_{\mu\nu}\delta_{\rho\sigma}+\delta_{\mu\rho}\delta_{\nu\sigma}+\delta_{\mu\sigma}\delta_{\rho\nu}).
\end{eqnarray}

According to the ansatz of the DS equation, the exponent of the ghost dressing function is given by $\kappa\sim \alpha_G$ and it is related to the exponent of the gluon dressing function $\alpha_D$ as $-2\kappa\sim \alpha_D$\cite{FN04}. 
When $\kappa=0.5$, corresponding to an infrared finite gluon propagator, the integral becomes
\begin{equation}
\delta^{bc}d(b,c)=-\frac{6.26379}{(p^2)^{0.5}}-(\frac{6.57974}{(p^2)^{0.5}}-2\frac{3.94784 p^2}{(p^2)^{0.5}}-\frac{1.12795 p^4}{(p^2)^{0.5}})
\end{equation}
Analysis of the exponent of the gluon dressing function of lattices with a long time axis is given in Ref.\cite{OS06}. 
In reality, $\kappa$ in the infrared and ultraviolet regions could be different, and 
the above numerical values should be regarded as simple estimations.  

The loop integral of the ghost-ghost-gluon triangle is given by
\begin{eqnarray}
&&I(p,a,b,c)=\int\frac{d^4 q}{(2\pi)^4}\left(\delta_{\mu\nu}-\frac{(p-q)_\mu(p-q)_\nu}{(p-q)^2}\right)\frac{Z_3((p-q)^2)}{(p-q)^2}\nonumber\\
&&\times\left(\frac{\delta^{a''c''}}{q^{2(1+\kappa)}}+2i f^{a''c''d}\phi_d(q)\right)(-g f^{a''aa'}q_\mu)\nonumber\\
&&\times\left(\frac{\delta^{b'a'}}{q^{2(1+\kappa)}}+2i f^{b'a'h}\phi_h(q)\right)(-g f^{b''bb'}q_\nu)\nonumber\\
&&\times\sum_x({\rm loop\quad term})(t^a)_{xx}.
\end{eqnarray}
where the quark loop contribution is expressed as the loop term.

\begin{table}[htb]
\begin{center}
\begin{tabular}{c c c c }
  $abc$ & $dh$ & $D_hD_dD_{Ae}$ & $D_{Ae}\phi_h\phi_d$  \\
\hline
 321 & 88 & 1.5 & -1.5 \\
     & 33 & 1.5 & -4.5  \\
     & 83 & 1.5 &  0   \\
     & 38 & 1.5 &  0  \\
\hline
 854 & 88 & 1.5 & -4.5  \\
     & 33 & 1.5 & -1.5  \\
     & 83 & 1.5 & $\sqrt 3/2$ \\
     & 38 & 1.5 & $\sqrt 3/2$ \\
\hline
 876 & 88 & 1.5 & -4.5  \\
     & 33 & 1.5 & -1.5 \\
     & 83 & 1.5 & $-\sqrt 3/2$\\
     & 38 & 1.5 & $-\sqrt 3/2$\\
\hline
\end{tabular}\label{color_matrix}
\caption{The $SU(3)$ color matrix elements of the ghost-ghost-gluon triangle.}
\end{center}
\end{table}

The color matrix elements are given in Table IV, %\ref{color_matrix}
where $D_{Ae}$ indicates the color diagonal gluon propagator.
 The coherent contribution, $D_e\phi\phi$ from $dh$=88 and 33 suggests that color
mixing occurs in the infrared region. The contribution of the color antisymmetric ghost propagator would be enhanced in an unquenched lattice simulation, 
in which case the quark loop becomes a source of the color antisymmetric pair.

The exponent of the color diagonal ghost dressing function $\alpha_G$ is $\sim 0.25$\cite{FN06a,FN04}, and that of the color antisymmetric ghost dressing function defined at $q\sim 0.4$ GeV is  $\alpha_G'\sim 0.9$\cite{FN06b}.  
By using the formulae in Ref.\cite{AFRvS03} 
and assuming  $Z_3((p-q)^2)$ is constant in the relevant integration region, we obtain
\begin{eqnarray}
&&I(p,a,b,c)\propto f^{abc}(c_1 K_2(2(1+\kappa),1,p)p^2+c_2 K_2(2(1+\kappa'),1,p)p^2)\nonumber\\
&&-\left( c_1 (K_2(2(1+\kappa),2,p)p^4-2 L_2(2(1+\kappa),2,p)p^4+M_3(2(1+\kappa),2,p)p^4\right)\nonumber\\
&&-\left(c_2 (K_2(2(1+\kappa'),2,p)p^4-2 L_2(2(1+\kappa'),2,p)p^4+M_3(2(1+\kappa',2,p)p^4\right)
\end{eqnarray}
where $c_1$ is the color matrix element of $D_{Ae}D_h D_d$, and $c_2$ is the
color matrix element of $D_{Ae}\phi_h \phi_d$.
Using $\kappa=0.5$ for $D_h, D_d$ and $\kappa'=0.9$ for $\phi_h, \phi_d$, 
we obtain
\begin{eqnarray}
&&I(p,a,b,c)\propto f^{abc}(c_1 (-\frac{\infty}{p^2})+c_2 (-\frac{6.1196}{p^{2.8}})\nonumber\\
&&-\left(c_1 (-\frac{\infty}{p^2}-2 \frac{\pi^2}{4p^2}+\frac{\pi^2}{16p^2})\right)\nonumber\\
&&-\left(c_2 (-\frac{2.203}{p^{3.8}}-2\frac{8.81215}{p^{3.8}}+\frac{2.03985}{p^{3.8}})\right).
\end{eqnarray}
We remark that the divergent terms cancel. 

The dressed ghost propagator can be incorporated into the ghost loop in the gluon propagator, as shown in Fig. 18, %\ref{gl_propg} 
and thereby modify the infrared behavior of the gluon propagator in the lattice
Landau gauge.
\begin{figure}[htb]
\parbox{\halftext}{
\centerline{\includegraphics[width=6.5 cm, height=3.9 cm]{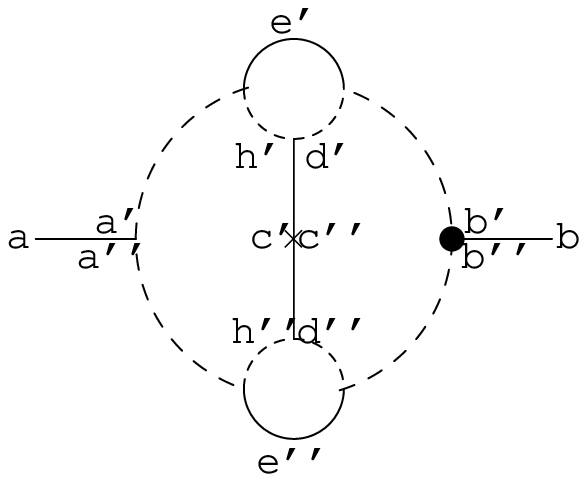}}
\caption{The gluon propagator with a ghost loop. The dashed line represents a ghost, the thin line a gluon. The $\times$ indicates dressing of the gluon in Cartan subalgebra.}}
\label{gl_propg}
\hfill
\parbox{\halftext}{
\centerline{\includegraphics[width=6.5 cm, height=3.9 cm]{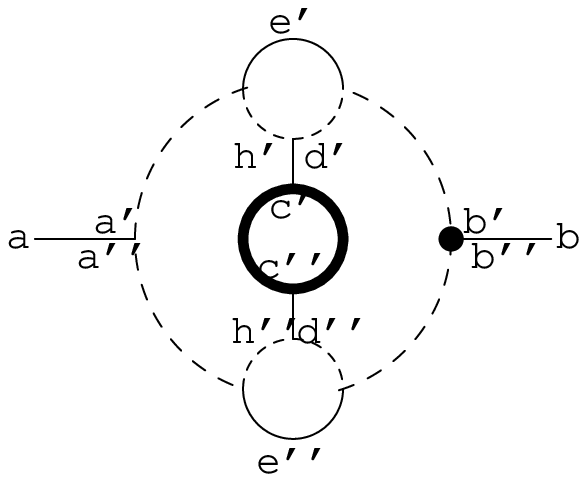}}
\caption{The gluon propagator with a ghost loop and quark loop contributions. The dashed line represents a ghost, the thin line a gluon, and the thick line a quark.}}
\label{gl_propg1}
\end{figure}

The loop of quarks in the fundamental representation yields ${\rm tr}(t^{c'}t^{c''})=\delta^{c',c''}\frac{1}{2}$. 
The sum of the color indices becomes 
\begin{equation}
2 g(a,b,c',c'')=\sum_{a',a'',b',b''}\delta^{c',c''}f^{aa'a''}f^{bb'b''}f^{c' a' b'}f^{c'' a'' b''}.
\end{equation}
This gives $4.5\delta^{ab}$.
The $D_{Ae}\phi_h\phi_d$ loop contribution, denoted by $c_2$, exists only when $c',c''=3,8$. There is no 
restriction to the $D_{Ae}D_hD_d$ loop contribution denoted by $c_1$, but the infrared singularity of the $c_2$ term is stronger than that of the $c_1$ term.

\begin{table}[tb]
\begin{center}
\begin{tabular}{ccccccccc}
c|ab &  11 & 22 & 33 & 44 & 55 & 66 & 77 & 88\\
\hline
3 & -0.25 & -0.25 & 2.25 & 0.5 & 0.5 & 0.5 & 0.5 &0.75\\
8 & 0.75 & 0.75 & 0.75 & 0 & 0 & 0 & 0 & 2.25\\
\hline
\end{tabular}
\caption{Color matrix elements of the gluon propagator.}
\end{center}
\end{table}

\section{The quark-gluon vertex}
The quark-gluon vertex is calculated from the longitudinal photon quark coupling
\begin{equation}
q_\mu\Gamma_\mu(p,q)=G(q^2)[(1-H(q,p+q))S^{-1}(p)-S^{-1}(1-H(q,p+q))],
\end{equation}
where $G(q^2)$ is the ghost dressing function and $H(q,p+q)$ is the
ghost-quark scattering kernel.
 In the continuum theory, the ghost-quark scattering kernel in the quark-gluon vertex is studied in Ref.\cite{Pa77}.  Its role in the DS approach is examined in Ref.\cite{FiAl03} and it is studied by lattice simulation in Ref.\cite{SkKi02}.

\begin{figure}[htb]
\centerline{\includegraphics[width=5 cm, height=6 cm]{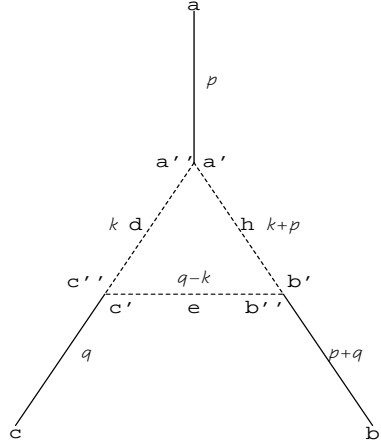}}
\caption{The ghost triangle. The color indices of the external gluons
and internal ghosts and the momentum assignments.}
\label{ghtriang}
\end{figure}

In the quark-gluon vertex, a ghost-triangle couples with an external gluon,
and two internal gluons couple to the quark, as shown in Fig.\ref{ghtriang}.

The color of the external gluon is specified by $a$, and the two internal gluons are
specified by $b$ and $c$.  When the ghost-quark scattering kernel is approximated by a dressed gluon exchange, the amplitude of the quark-gluon vertex becomes
\begin{eqnarray}
&&M(p,s,a,x,y)=\int\frac{d^4 k}{(2\pi)^4}\int\frac{d^4 q}{(2\pi)^4}\nonumber\\
&&\times \left( \frac{\delta^{a''c''} }{k^{2(1+\kappa)}} + 2i f^{a''c''d}\phi_d(k)\right) (-g f^{a''aa'}k)\nonumber\\
&&\times \left(\frac{\delta^{c'b''}}{(q-k)^{2(1+\kappa)}}+2i f^{c'b''e}\phi_e(q-k)\right) \left(-g f^{c''cc'}(q-k)\right)\nonumber\\
&&\times\left(\frac{\delta^{b'a'}}{(k+p)^{2(1+\kappa)}}+2i f^{b'a'h}\phi_h(k+p)\right )\left(-g f^{b''bb'}(k+p)\right)\nonumber\\
&&\times(ig t^c)_{xz}(i g t^b)_{zy} Z_2((s-q)^2)\frac{-i(\Slash{s}-\Slash{q})+M}{(s-q)^2+M^2}\nonumber\\
&&\times(\delta_{\mu\nu}-\frac{q_\mu q_\nu}{q^2})\frac{Z_3(q^2)}{q^2}\nonumber\\
&&\times(\delta_{\eta\delta}-\frac{(q+p)_\eta (q+p)_\delta}{(q+p)^2})\frac{Z_3((q+p)^2)}{(q+p)^2}.
\end{eqnarray}

It is important to note that the color antisymmetric ghost does
not appear as an external line, but, rather, it appears in the vertex as an internal
line. The gluon dressing indicated by the cross can be contracted as $\delta^{c'c''}$ and becomes a propagator in the Cartan subalgebra. This could effectively
introduce the ghost condensates effect, i.e. a choice of a specific color direction.

\section{Effects on the Kugo-Ojima color confinement parameter and the running coupling}
 Kugo and Ojima\cite{KO79} constructed a two-point function $u^{ab}(q)$ from the Lagrangian in the Landau gauge that satisfies BRST symmetry. They claimed that if its value at momentum zero is $-1$, it is evidence of color confinement: 
 
\begin{eqnarray}
(\delta_{\mu\nu}-\frac{q_\mu q_\nu}{q^2})u^{ab}(q^2)&=&\frac{1}{V}
\sum_{x,y} e^{-ip(x-y)}\langle  {\rm tr}\left({\Lambda^a}^{\dag}D_\mu \displaystyle{\mathcal M}^{-1}[A_\nu,\Lambda^b] \right)_{xy}\rangle,\nonumber\\
u^{ab}(0)&=&-\delta^{ab}c.
\end{eqnarray}
In this argument and in the proof of the Slavnov-Taylor identity regarding the ratio of the vertex renormalization factor and the wave function renormalization factor,\cite{Kugo95}
\[
\frac{Z_1}{Z_3}=\frac{\tilde Z_1}{\tilde Z_3}=\frac{Z_{\bar\psi\psi A}}{Z_\psi},
\]
the ghost propagator is assumed to be color diagonal.

The replacement
\begin{equation}
\langle c^a(x)\bar c^b(y)\rangle=D_G(x-y)\delta^{ab}+f^{abc}\phi^c(x-y)
\end{equation}
does not affect the argument of the tree level, since the expectation value of $\phi(x-y)$ is 0. However, when there is a ghost color mixing of the type depicted in Fig. 15, %\ref{gh_propg}
we have
\begin{equation}
\langle c (A_\nu\times \bar c)\rangle_{1PI}\ne iq_\rho\langle (A_\rho\times c)(A_\nu\times \bar c)\rangle_{1PI}.
\end{equation}
Thus, when we define
\[
\langle c\bar c\rangle\equiv -\frac{1}{q^2 G(q^2)},
\]
\[
\langle (A_\mu\times c)\bar c\rangle_{1PI}\equiv -iq_\mu F(q^2)
\]
and 
\begin{equation}
\langle D_\mu c\bar c\rangle=\langle \partial_\mu c\bar c\rangle+\langle (A_\mu\times c)\bar c\rangle\equiv iq_\mu(1+F(q^2))\frac{1}{q^2 G(q^2)},\label{ko}
\end{equation}
we obtain
\begin{eqnarray}
\langle D_\mu c(A_\mu\times \bar c)\rangle&=&\langle \partial_\mu c\bar c\rangle\langle c(A_\mu\times \bar c)\rangle_{1PI}+\langle (A_\mu\times c)(A_\nu\times \bar c)\rangle_{1PI}\nonumber\\
&\ne& (\delta_{\rho\mu}-\frac{q_\mu q_\rho}{q^2})\langle c(A_\mu\times \bar c)\rangle_{1PI}\label{1pieq}.
\end{eqnarray}
In other words, although the lattice simulation confirms $G(0)=0$\cite{FN04}, there should be a contribution of the ghost propagator of the type appearing in Fig. 15 %\ref{gh_propg} 
between $c$ and $\bar c$ that is not proportional to $\delta^{ab}$, and $1+u(0)=1+F(0)$ in eq.(\ref{ko}) is not necessarily equal to $G(0)=0$. 

In the $S$-matrix theory, the ghost intermediate state cancels the intermediate states with a non-physically polarized gluon.  The longitudinal gluon with
polarization vector $-ip_\mu$ couples with
 a set of diagrams of a given order in $g$ and a given number of transverse gluons on mass shell, which are expressed as circles in the t' Hooft definition, and  it was shown that all the contributions cancel, yielding 0, when the ghosts are color diagonal.

 Our analysis suggests that if the longitudinal gluon forms a pair of color antisymmetric ghosts and the loop intersects with the circle, the contribution
 remains since the contribution of the cutoff denoted by $\Lambda$ in the
 t' Hooft definition cancels the color-diagonal ghost contribution.
This results in an artificial suppression of the ghost propagator and the QCD running coupling.

No infrared suppression occurs in $\alpha_I({\Vec q})$,
since the longitudinal gluon polarized in the 4th direction does not contribute
to the three-dimensional calculation. 

The color indices of the color antisymmetric ghost pair could be ordered
if there is a long-range background gluon field that couples to the pair. In this case, Eq.(\ref{1pieq})
would be satisfied and the Kugo-Ojima parameter $c$ would become close to 1.

There is no direct proof that the quark loop introduces ordering, but the difference between the Binder cumulants of the color antisymmetric ghost propagator of the quenched simulation and the unquenched simulations\cite{FN06a,FN06b,FN06d} strongly
suggests this effect. The temperature dependence of the parameter $c$\cite{FN06d} could also
 be explained by this mechanism.

\section{Discussion and outlook}
We investigated whether the deviation of the Kugo-Ojima parameter $c$ from 1
and the infrared suppression of
the effective coupling $\alpha_s(q)$ in the Landau gauge could be due to the
color antisymmetric ghost propagator.

In the case of the Coulomb gauge, the ghost dressing function is three dimensional,
 no dissipation of the flow occurs and the running coupling freezes.

The lattice Landau gauge results show some discrepancies from the results of the
DS equation with regard to the infrared exponents of the ghost propagator
and the gluon propagator. We investigated the possibility that this discrepancy
comes from the color-mixing of the ghost propagator induced by the 
ghost-gluon-ghost triangle diagram. A study employing the DSE is left as a future project.

The color antisymmetric ghost pair couples to a gluon whose color
is in the Cartan subalgebra. The lattice data suggest that the
 gluon in the Cartan subalgebra couples with a dynamical quark loop,
 and this introduces a difference between the color antisymmetric ghost propagator of the
quenched configurations and that of the unquenched configurations.

We also studied the contribution of ghost triangle diagram in the quark-gluon
vertex. For the estimation of the latter effect, it is necessary to perform a two-loop calculation, and this is left to a future.

\section*{Acknowledgements}
The main part of this work was done at the Department of Theoretical Physics of University of Graz in August 2007. The author thanks Kai Schwenzer and Reinhard Alkofer for helpful and enlightening discussions in Graz and Hideo Nakajima for the
collaboration in the lattice simulation cited in this paper.
Thanks are also due to the Austrian Academic Exchange Service for their support during the author's stay in Graz and the Japan Society for the Promotion of Science (JSPS) for support through the Scientist Exchange Program that enabled the collaboration.

%\appendix
%\section{First Appendix} %Empty argument \section{} yields `Appendix'. 
%
%\section{Second Appendix}

\end{document}